\documentclass[11pt]{article}
\usepackage{graphicx}  
\newcommand{\greeksym}[1]{{\usefont{U}{psy}{m}{n}#1}}
\newcommand{\umu}{\mbox{\greeksym{m}}}

\newcommand\lsim{\mathrel{\rlap{\lower4pt\hbox{\hskip1pt$\sim$}}
    \raise1pt\hbox{$<$}}}
\newcommand\gsim{\mathrel{\rlap{\lower4pt\hbox{\hskip1pt$\sim$}}
    \raise1pt\hbox{$>$}}}

\date{}

\begin{document}

\title{\bf\Large Models of inflation, supersymmetry breaking \protect\newline 
and observational constraints}
%
%
%
%
%
\author{Laura Covi\\
DESY - Theory Group, Notkestrasse 85,\\ 
D-22603 Hamburg, Germany
}
%
%
%

\maketitle              

\begin{abstract}
We review the connection between inflationary models and observations and 
concentrate to describe models based on softly broken supersymmetry, in 
particular running mass models, and their predictions. We then present 
a fit of the spectral index of the curvature perturbation, 
assuming a flat $\Lambda$CDM cosmology. 
\end{abstract}

\section{Introduction}

An epoch of inflationary expansion of the Universe is necessary to 
solve some of the problems of the standard Big Bang cosmology, like the 
large scale homogeneity and isotropy, flatness and unwanted relics problems.
Moreover it has been also found that slow roll inflation can generate 
the small scale structure in the Universe through a primordial gaussian
curvature perturbation, originated from the quantum fluctuations of
the inflaton field.

In general a model of inflation consists in a scalar potential 
$V(\phi)$ for the inflaton field $\phi $ satisfying slow roll 
conditions\footnote{As usual, in the following 
$M_P=2.4\times 10^{18}\mbox{GeV}$ is the Planck mass, 
$a$ is the scale factor and $H=\dot a/a$ is the Hubble parameter, and
$k/a$ is the wavenumber.} \cite{lr99}:
\begin{eqnarray}
\varepsilon & = & M_P^2\left({V''\over V}\right)^2 \ll 1\\
\eta        & = & M_P^2\left({V'\over V}\right)^2 \ll 1
\label{slowroll}
\end{eqnarray}
where the prime denotes derivative with respect to the field $\phi$.

The point of contact between observation and models of inflation
is the  spectrum ${\cal P}_{\cal R}(k)$ of the curvature
perturbation, which, in the slow roll approximation $3H\dot\phi\simeq -V'$, 
 is given in terms of the inflaton potential $V(\phi)$ by
\begin{equation}
{4\over 25}{\cal P}_{\cal R}(k)
 = \frac{1}{75\pi^2 M_P^6}\frac{V^3}{V'^2} |_{k=aH} \,,
\label{delh}
\end{equation}
where the potential and its derivatives are evaluated at the epoch of 
horizon exit
$k=aH$. To work out the value of $\phi$ at this epoch one uses
the relation
\begin{equation}
\ln(k_{\mathrm end}/k)\equiv N(k)
=M_P^{-2}\int^\phi_{\phi_{\mathrm end}} (V/V') d\phi
\,,
\label{Nofv}
\end{equation}
 where $N(k)$ is actually the number of $e$-folds from horizon exit of
the scale $k$ to the end of slow-roll inflation.
At the scale explored by the COBE measurement of the cosmic microwave
background (cmb) anisotropy, 
 $N(k_{\mathrm COBE})$ depends on the expansion of the Universe after inflation
in the manner specified by:
\begin{equation}
N_{\mathrm COBE} \simeq 60 - \ln(10^{16}\, \mbox{GeV}/V^{1/4}) - \frac{1}{3}
\ln(V^{1/4}/T_{\mbox{reh}}) 
\, .
\label{Ncobe}
\end{equation}
In this expression, $T_{\mbox{reh}}$ is the  reheat temperature, and
instant reheating is assumed. 

Given the above relations, the observed large-scale normalization  
${\cal P}_{\cal R}^{1/2}\simeq 10^{-5}$ provides a strong 
constraint on models of inflation.
But here we are interested in the scale-dependence of
the spectrum, defined by the, in general, scale-dependent spectral 
index $n$;
\begin{equation}
n(k)-1\equiv \frac{d \ln {\cal P}_{\cal R}(k)}{ d \ln k}
\,.
\end{equation}
According to most inflationary models, $n$ has negligible variation on
cosmological scales so that ${\cal P}_{\cal R} (k)\propto k^{n-1}$, but
we shall also discuss an interesting class of models giving
a different scale-dependence.

From (\ref{delh}) and (\ref{Nofv}),
\begin{eqnarray}
n-1 &=&  2 M_P^2 (V''/V)-3 M_P^2 (V'/V)^2 
\,,
\label{nofv}
\end{eqnarray}
and in  almost all models of inflation, (\ref{nofv}) 
is well approximated by
\begin{equation}
n-1=2 M_P^2(V''/V)
\label{nofvapprox}
\,.
\end{equation}
We see that the  spectral index 
 measures the {\em shape} of the inflaton potential $V(\phi)$,
being independent of its overall normalization. For this reason, 
it is a powerful discriminator between models of inflation.

\section{Supersymmetric models of hybrid inflation}

A very successful and interesting type of inflationary model
is hybrid inflation, proposed by Linde \cite{Linde93}.
In this case the potential depends on two fields, the inflaton
and another field, which triggers a phase transition and
the end of slow roll inflation and acquires a non vanishing 
v.e.v. at the end.
The basic potential is of the form:
\begin{equation}
V = V_\phi (\phi) + \frac{1}{2} \left(\phi^2-m^2_\psi\right) \psi^2 +....
\end{equation}
so, for $\phi \gg m_\psi$, the hybrid field $\psi $ has vanishing
v.e.v. and inflation can take place driven by $V_\phi$, but when
the inflaton becomes smaller, $\psi$ is displaced from the origin
and its v.e.v. stops slow roll inflation.
Then a phase of oscillation of the two fields around the true
minimum of the potential occurs and the preheating/reheating process.

We see then that the characteristic of hybrid inflation is that
two different part of the potential are responsible for the inflationary
phase and for the following stage of reheating. So the power spectrum
of the curvature perturbations depends only on the inflationary
potential $V_\phi$ and the critical value $\phi_c = m_\psi $, which
corresponds to the end of slow roll inflation.

Supersymmetric versions of hybrid inflation have been studied
by many authors (see \cite{lr99} and references therein) and are 
based on the superpotential
\begin{equation}
W = \lambda \Phi (M^2-\Psi\bar\Psi),
\end{equation}
where capital Greek letter denote the superfields corresponding
to the lower case Greek letter scalar field.

In the limit of global supersymmetry, such potential is perfectly 
flat with respect to the inflaton field and the small slope needed
for slow roll is generated by supersymmetry breaking effects.
Depending on the different mechanism considered, different scenarios
are possible, as described in Table \ref{LC-Tab1}, together with the
prediction for the spectral index and its derivative.

\begin{table}
\caption{Various models of hybrid inflation and their prediction
for the spectral index as a function of $N= -\log(k/k_{\mathrm end})$
and its derivative, assuming small field values. 
In the case of the simple linear term, $n-1$ is given by the full
expression (\ref{nofv}) since $\eta $ vanishes, but note that
in supergravity generally, other contribution to the spectral index
coming from higher order terms are usually present and can be larger
than the one listed here \cite{bcd00}.}
\begin{center}
\renewcommand{\arraystretch}{1.4}
\setlength\tabcolsep{15pt}
\begin{tabular}{@{}lllp{2.5cm}}
\hline\noalign{\smallskip}
$V_\phi(\phi)/V_0$ & $n-1$ & $d n/d\log(k)$ & Origin of the slope \\
\noalign{\smallskip}
\hline
\noalign{\smallskip}
$1-\frac{\lambda^2}{4\pi^2} \log\left({\sqrt{2}\lambda\phi\over Q}\right)$ 
& ${-1\over N+\frac{2\pi^2\phi_c^2}{\lambda^2 M_P^2}}$ &
$ {-1\over\left(N+\frac{2\pi^2\phi_c^2}{\lambda^2 M_P^2}\right)^2}$ &
1 loop for spont. broken susy\\
\noalign{\smallskip}
$ 1\pm \frac{1}{2} \frac{m^2_\phi}{V_0} \phi^2$ & 
$\pm \frac{m^2_\phi M_P^2}{V_0}$ & $0$
& Susy breaking mass \\
\noalign{\smallskip}
$1 \pm \beta \frac{\phi}{M_P} $& $- 3\beta^2$ & $0$ &
Susy breaking linear term\\
\noalign{\smallskip}
$1 \pm \frac{\phi^4}{M_P^4} $ & ${12\over \frac{M_P^2}{2 \phi_c^2}\mp N}$&
  ${\mp 12\over \left(\frac{M_P^2}{2 \phi_c^2}\mp N\right)^2}$ &
Sugra quartic term\\
\noalign{\smallskip}
\hline
\noalign{\smallskip}
\end{tabular}
\end{center}
\label{LC-Tab1}
\end{table}

Note that the COBE normalization constraints the overall scale $V_0$,
while the spectral index gives direct informations on the supersymmetry
breaking effect which dominates the inflaton potential.

\newpage

\section{Running mass models}

\subsection{The potential}

Let us consider now a specific type of models of hybrid inflation,
the models with a running inflaton mass\cite{st97,cl98,clr98,c98,rs}. 
In these models, based on softly broken supersymmetry\footnote{
We denote as spontaneously broken supersymmetry any theory where
supersymmetry breaking preserves the super-trace relation 
${\cal S}\mbox{tr} {\cal M}^2 = 0$, while in softly broken 
supersymmetry one has
${\cal S}\mbox{tr} {\cal M}^2 \neq 0$; e.g. the displacement of the inflaton 
field from the minimum of the potential causes a spontaneous breaking 
in the inflaton sector (which can give a soft breaking in the visible
sector), while spontaneous or explicit supersymmetry breaking in
another (hidden) sector is needed to have soft breaking in the inflaton
sector. Note that in the first case, supersymmetry is restored once
the inflaton field settles in the true minimum, while in the second
case the breaking persists all the time, even if it can be affected
by the dynamics during inflation.
},
one-loop corrections to the tree-level potential are taken into
account by evaluating the inflaton mass-squared
 $m^2(\ln (Q))$ at the renormalization 
scale $Q\simeq \phi$, as long as $\phi$ greater than all other 
relevant scales. 
\begin{equation}
V=V_0 + {1\over 2} m^2(\ln(Q)) \phi^2 + \cdots
\,.
\label{runpot}
\end{equation}
The dependence on $Q$ is only logarithmic due to supersymmetry, which
ensures the cancelation of the quadratic divergences; anyway since 
supersymmetry is broken by the soft terms, a logarithmic
divergence survives.

Over any  small  range of $\phi$, 
it is a good approximation to  take the running mass to be  a 
linear function of $\ln\phi$ and then one reproduces explicitly
the expression of the tree potential plus loop correction:
\begin{equation}
V=V_0 + {1\over 2} m^2(\ln(Q)) \phi^2 - {1\over 2} c (\ln(Q)) 
{V_0\over M_P^2} \phi^2 \ln (\phi/Q) + \cdots
\,.
\label{vlin}
\end{equation}

It has been shown \cite{cl98} that the linear approximation
is very good over the range of $\phi$ corresponding to horizon exit
for scales between $k_{COBE}$ and $8h^{-1}{\mathrm Mpc}$. 
We shall want to estimate the reionization epoch, which involves a
 scale of order $k_{reion}^{-1}\sim
10^{-2}{\mathrm Mpc}$ (enclosing the relevant mass of order
$10^6\odot$). Since only a crude estimate of the reionization
epoch is needed, we shall assume that the linear approximation is
adequate down to this `reionization scale'. 

A potential of the type (\ref{vlin}) gives rise to four different
models of inflation, depending on the sign of $c$ and the
direction of rolling of the inflaton field, towards or away from the
origin\footnote{We follow the labeling introduced in
\cite{cl98}.}. 

The value of $c$ is given by the well-known Renormalization 
Group Equation for the inflaton mass, and it depends on the 
gauge and Yukawa couplings, $\alpha $ and $\lambda $ respectively, 
of the inflaton field:
\begin{equation}
\frac{V_0 c}{M_P^2} = - {d m^2\over d t} =  
\frac{2 C}\pi \alpha \widetilde m^2 
- {D\over 16\pi^2} |\lambda |^2 m^2_{loop}
\,.
\label{c-def}
\end{equation}
Here, $C$ is a positive  group-theoretic number of order 1,  
 $\widetilde m$ is the supersymmetry breaking gaugino mass,
$D$  is  a positive constant counting the number of 
scalar particles interacting with the inflaton through $\lambda $
and $m^2_{loop}$ is their common susy breaking mass-squared.
Negligible supersymmetry breaking trilinear coupling is assumed
in the above expression.

We see that the running is then directly connected to the
couplings of the inflaton field and the supersymmetry breaking
mass spectrum of our model and so to complete our estimate of $c$, 
we need  the gaugino or scalar mass. 
A very minimal and traditional hypothesis is that soft supersymmetry 
breaking is gravity-mediated and that the scale of susy breaking
during inflation $V_0^{1/4}$ coincides with the scale of supersymmetry
breaking in the present vacuum, $M_S\equiv \sqrt F$, 
where $F$ is the auxiliary field responsible for spontaneous 
supersymmetry breaking  in the hidden sector.

With gravity-mediated susy breaking, typical values of the 
 masses are $\widetilde m^2\sim |m_{loop}^2|\sim V_0/M_P^2$,
which makes $|c|$ of order of the coupling strength
$\alpha$ or  $|\lambda|^2$. 

At least in the case of dominance of the gauge coupling, one
then expects a small, but non--negligible value for $c$
\begin{equation}
|c|\sim 10^{-1}\mbox{ to }10^{-2}
\,.
\label{crange}
\end{equation}
In special versions of gravity-mediated susy breaking,
 the masses could be much smaller, leading to $|c|\ll 1$. In that case,
the mass would hardly run, and the spectral index would be practically
scale-independent. With gauge-mediated
susy breaking, the masses  could be much bigger; this would not
lead to a model of inflation
(unless the coupling is very suppressed) because
it would not satisfy the  slow-roll requirement
$|c|\lsim 1$. 

So we see that both from the theoretical and from the observational
perspective, the interesting parameter region for the running mass
models is given by (\ref{crange}).

\subsection{The spectrum and the  spectral index}

Using (\ref{Nofv})  we find
\begin{eqnarray}
s e^{c\Delta N(k)} &=& c \ln(\phi_*/\phi)\label{sigma}\\
\Delta N(k)&\equiv& N_{COBE} - N(k)
\equiv \ln(k/k_{COBE})
\,,
\end{eqnarray}
where 
$s$ is an integration constant, which absorbs the dependence
on the end of inflation. The spectral index is then given,
using (\ref{nofvapprox}), by 
\begin{equation}
{n(k)-1\over 2} = 
s e^{c\Delta N(k)} - c 
\label{runpred}
\,.
\end{equation}
We see then that a relatively strong scale dependence
arises for $n(k)$, depending on the magnitude of the
constants $s$ and $c$. To satisfy the slow-roll 
conditions (1), both $c$ and $s$ 
must be smaller than 1 in magnitude.

In order to evaluate the power spectrum later on (see (\ref{peakpresc})
in the following section), we also need the variation of
$\delta_H$ which comes from integrating this expression,
\begin{equation}
\frac{\delta_H(k)}{\delta_H(k_{COBE})}
=\exp\left[ \frac sc \left(e^{c\Delta N}-1\right)-c\Delta N \right]
\,.
\label{delta_H}
\end{equation}

We are mostly interested in cosmological scales between $k_{COBE}$
and $k_8$, corresponding to $0\lsim \Delta N\lsim 4$.
In this range the scale-dependence of $n$ is approximately linear
(taking $|c|\lsim 1$)
and the variation $\Delta n\equiv n_8-n_{COBE}$ is given approximately
by
\begin{equation}
\Delta n \simeq 4
\frac{d n}{d \ln k}\simeq 8sc
\,.
\label{scaledep3}
\end{equation}

\section{Fit to a
$\Lambda$CDM model}

We will present here a global fit to the cosmological parameters
according to the procedure described in full detail in \cite{lc00}.
Following \cite{cl00} we will take into account also the recent
measurements of the cmb anisotropy performed by the Boomerang 
\cite{boom} and Maxima-1 \cite{maxima} balloon experiments.

The observational constraints on the cosmological parameters and
on the spectral index have been studied by many authors, but 
in our analysis we will use a different treatment of reionization
and we will compare the present data with
the two parameters prediction for 
the scale dependence of the  spectral index in running mass models.

Observations of various types  indicate that we live in a low density 
Universe, which is at least approximately flat (see e.g.
\cite{boom,maxima}). In the interest of  simplicity
we therefore adopt the  $\Lambda$CDM model, defined by the
requirements that the Universe is exactly flat, and that the 
non-baryonic dark matter is cold with negligible interaction.
Essentially  exact flatness is predicted by inflation, unless one invokes
a special kind of model, or special initial conditions.
We shall constrain the parameters of the $\Lambda$CDM model, including
the spectral index, by performing a least-squares fit to 
 key observational quantities.

\subsection{The parameters}
The $\Lambda$CDM model is defined by the spectrum
${\cal P}_{\cal R}(k)$  of the
primordial curvature perturbation, and 
 the four   parameters that are
needed to translate this spectrum into spectra for
the  matter density perturbation and the cmb anisotropy.
The   four parameters are the
 Hubble constant $h$ (in units of 
$100 {\mathrm km s}^{-1}{\mathrm Mpc}^{-1}$),
 the total matter density parameter $\Omega_0$,  the 
baryon density parameter $\Omega_b$, and the 
reionization redshift $z_ R$. As we shall describe, 
 $z_ R$ is  estimated by assuming that reionization occurs when some
fixed fraction $f$ of the matter collapses.
Within the reasonable range $f\sim 10^{-4}$ to $1$, the main results
are insensitive to the precise value of $f$.

The spectrum is conveniently  specified by its value at a scale explored
by COBE, and the spectral index $n(k)$.
We  shall  consider the usual case of  a constant spectral index,
and the case of running mass models where $n(k)$  is given by
the two-parameter expression in (\ref{runpred}).  
Since ${\cal P}_{\cal R}(k_{\mathrm COBE})$ is determined
very accurately by the COBE data we fix its value.
Excluding   $z_ R$ and ${\cal P}_{\cal R}(k_{COBE})$,
  the $\Lambda$CDM model is  specified by
four  parameters in the case of a constant spectral index, or by five
parameters in the case of running mass inflation models.

\subsection{The data}

As described in detail in \cite{lc00,cl00}, we consider a sample of
data constituted by seven  observational quantities.
Of these quantities, three are
 the cosmological  quantities $h$, $\Omega_0$, $\Omega_ B$, which  
we are also taking as free
parameters. We will assume that, at least at some
crude level, we can pretend that the errors are all
random and uncorrelated, and perform a least squares fit.

The other data are taken so to sample different observable scales:
one one side we use large scale structure observation, on
the other measurements of the cmb anisotropy.

So, first of all we consider the rms density perturbation at 
$8h^{-1}\mbox{Mpc}$, $\sigma_8$, measured through the 
abundance of rich galaxy clusters at redshift $z=0$ to a few
\cite{vl} and the shape parameter $\Gamma $ that specifies
the slope of the galaxy correlation function on  
scales of order $1h^{-1}$ to $100h^{-1}{\mathrm Mpc}$
\cite{llvw,will}

Secondly we take also two data from the latest measurements  
of the cmb anisotropy, i.e. the height of the first peak
at $\ell \simeq 210-30 $ and the ratio between the heights
of the second and first peak. We consider the average of
the two experiment \cite{boom,maxima} and 
include the calibration error in quadrature.

We fix the value of the COBE normalization, as described
in detail in \cite{lc00}.

The adopted values and errors are given in Table 2 and 3, with the
results of the fit for a constant  and scale--dependent $n$. 

\subsection{Reionization}

The effect of reionization on the cmb anisotropy is determined by  the
optical depth $\tau$. We assume  sudden, complete reionization
at redshift  $z_ R$, so that the optical depth $\tau$ is given by
\cite{peacock}
\begin{equation}
\tau= 0.035\frac{\Omega_b}{\Omega_0} h \left( \sqrt{\Omega_0(1+z_ R)^3 +1
-\Omega_0} -1 \right)
\,.
\end{equation}

In previous investigations,  $z_ R$ has been regarded as a free 
parameter, usually  fixed at zero or some other value.  In this 
investigation, we instead estimate $z_ R$, in terms of the parameters 
that we are varying plus assumed astrophysics. Indeed, it is usually 
supposed that reionization occurs at an early epoch, when some fraction 
$f$ of the matter has collapsed into objects with mass very roughly  
$M=10^6_\odot$. Estimates of $f$ are in the range \cite{llreion}
\begin{equation}
10^{-4.4}\lsim f\lsim 1
\label{fest}
\,.
\end{equation}
In the case $f\ll 1$, the 
 Press-Schechter approximation gives the estimate 
\begin{equation}
1+z_ R \simeq\frac{\sqrt2 \sigma(M)}{\delta_ c g(\Omega_0)}
{\,\rm erfc}^{-1}(f)
\hspace{4em}(f\ll 1)
\label{fll1}
\,.
\end{equation}
Here  $\sigma(M)$ is the present, linearly evolved,
 rms density contrast  with   top-hat smoothing, 
 and $\delta_ c=1.7$ is the overdensity required
for gravitational collapse~($g$ is the suppression 
factor of the linearly
evolved density contrast at the present epoch, which does not apply
at the epoch of reionization. See \cite{lc00} for details).
In the case $f\sim 1$,  one can justify only the 
rough estimate
\begin{equation}
1+z_ R \sim \frac{ \sigma(M)}{g(\Omega_0)}
\hspace{4em}(f\sim 1)
\, ,
\end{equation}
not very different from the one that would be obtained
by using $f=1$ in (\ref{fll1}).

In our fits, we fix $f$ at different values in the above range,
and find that the most important results are not very sensitive to $f$
even though the corresponding values of $z_R$ can be quite different.

\subsection{The predicted peak height}

The  CMBfast package \cite{CMBfast} gives $C_\ell$, for given values 
of the parameters with $n$ taken to be scale-independent.
Following  \cite{martin}, we parameterize the 
CMBfast output at the first peak   in the form
\begin{equation}
\widetilde C_{peak} = \widetilde C_{peak}^{(0)} 
\left(\frac{220}{10}\right)^{\nu/2}
\,,
\label{eq:cpeak}
\end{equation}
where $\widetilde C_{peak}^{(0)}$ is the value of $\widetilde C_{peak}$
at the reference value of the parameters, where $\nu = 0$, and 
\begin{equation}
\nu \equiv a_n(n-1)
 + a_h \ln(h/0.65) + a_0\ln(\Omega_0/0.35)
+a_b h^2(\Omega_b - 0.019) -0.65f(\tau)\tau
\,.
\end{equation}

The coefficients are $a_n=0.88$, $a_h= -0.37$, $a_0=-0.16$,
$a_b= 5.4$, and $\widetilde C_{peak}^{(0)}=77.5\umu {\mathrm K}$. 
The formula reproduces the CMBfast results within 10\% for a 1-$\sigma$
variation of the cosmological parameters, $h, \Omega_0$ and $ \Omega_b$,
and $n_{\mathrm COBE}=1.0\pm 0.05$.  

With the function $f(\tau)$ set equal to 1, the term $-0.65 \tau$ is 
equivalent to multiplying $\widetilde C_{peak}$ by the usual factor 
$\exp(-\tau)$. 
We use the following formula, which 
was obtained by fitting the output of CMBfast, and is
accurate to a few percent over the interesting range of $\tau$;
\begin{equation}
f=1- 0.165 \tau/(0.4+\tau)
\,.
\end{equation}

For the running-mass model, we start with  the above estimate for $n=1$,
and rescale it according to the scale dependence of (\ref{delta_H}), i.e. 
\begin{equation}
\frac{\widetilde C_{peak}}{
\sqrt{\widetilde C_{peak}^{(n=1)}}
} = \frac{\delta_H(k(\ell,\Omega_0))}{\delta_H(k_{\mathrm COBE}(\Omega_0))}
\,.
\label{peakpresc}
\end{equation}
In the case of constant $n$, this  prescription
 corresponds to the previous one with $a_n=0.91$, in good agreement
with the output of CMBfast.

\section{Results}

\subsection{Constant $n$}

For the case of a constant spectral index our results are summarized
in Table (\ref{lc-table2}), together with the data points used. 

\begin{table}
\caption{Fit  of the $\Lambda$CDM model to presently available data,
assuming reionization when a fraction $f=10^{-2.2}$ of matter
has collapsed (corresponding redshift at best fit is $z_R=18$).
The first two rows show the data points, while the result of the  
least-squares fit for the parameters is given 
in the lines three to five.  All uncertainties are at the nominal 1-$\sigma$
level. The total $\chi^2$ is 6.3  with  three  degrees of freedom}
\begin{center}
\renewcommand{\arraystretch}{1.4}
\setlength\tabcolsep{7pt}
\begin{tabular}{@{}lllllllll}
\hline\noalign{\smallskip}
& $n$ & $\Omega_b h^2$ & $\Omega_0$ & $h$ 
&$\widetilde \Gamma$ & $\widetilde \sigma_8$ & 
$\sqrt{\widetilde C_\ell^{1st}}$ &
${\widetilde C_\ell^{2nd}\over\widetilde C_\ell^{1st}}$\\
\noalign{\smallskip}
\hline
\noalign{\smallskip}
data & --- & 0.019 & 0.35 & 0.65 &
 0.23 & 0.56 & $74.0\,\umu K$ & 0.38 \\
err. & --- & 0.002 & 0.075 & 0.075  & 0.035 & 0.059 & $5.0\, \umu K$
& $0.06$
 \\
fit & $0.99$ & $0.021$ & $0.38$ & $0.62$
 & 0.19  & 0.56  & $70.8\, \umu K$ & 0.49\\
err. & 0.05 & 0.002 & 0.06 & 0.05  & --- & --- & --- & --- \\
$\chi^2$ & --- & 0.9  & 0.2 & 0.2 &
 $1.3$ & $0.002$ & $0.4$ & 3.3 \\
\noalign{\smallskip}
\hline
\end{tabular}
\end{center}
\label{lc-table2}
\end{table}

The relative high value of the $\chi^2$ is mainly due to the second peak
contribution, but note that the discrepancy between the measured and
fitted values of the peak ratios is within the $2\sigma $ limit and that
the best fit value of $\Omega_b h^2 $ is also within $1\sigma $ from the
Big Bang Nucleosynthesis measurement. 

We show also in Figure 1, the comparison of the dependence of the 
best fit value of the spectral index on the the reionization redshift 
with the dependence on the reionization fraction $f$. Note that a
strong correlation between $n$ and $z_R$ is present, while the
correlation with $f$ is much weaker, and in particular the lower
bound on $n$ is practically independent of $f$. Low values of $z_R$ 
and therefore high $f$ have lower $\chi^2$.

\begin{figure}[t]
\begin{center}
\includegraphics[width=.6\textwidth]{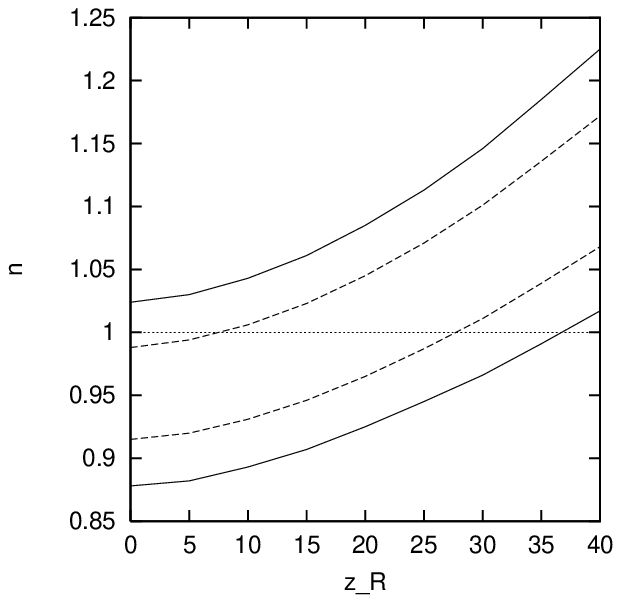}
\includegraphics[width=.6\textwidth]{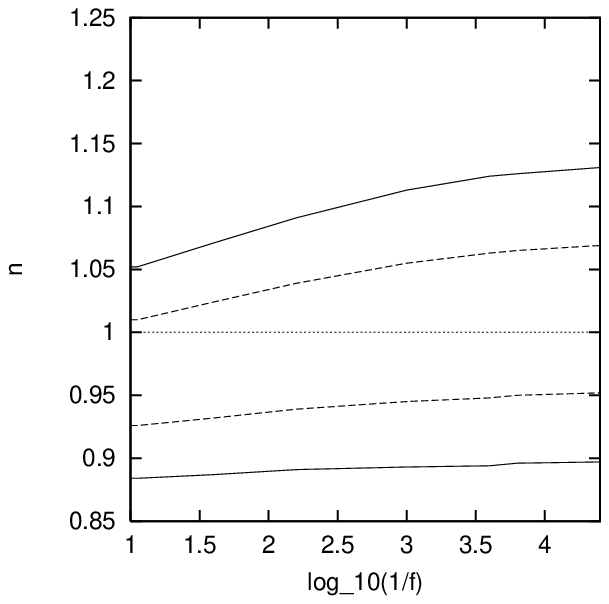}
\end{center}
\caption{The plots show nominal 1- and 2-$\sigma$ bounds on $n$. 
In (a) the fit is performed fixing the reionization epoch $z_R$, while
in (b) is fixed instead the fraction  $f$ of matter which has collapsed 
at the epoch of reionization (the corresponding reionization redshifts, 
at best fit, are in the range $10$ to $26$)}
\label{fig1}
\end{figure}

\subsection{Running mass models}

We performed the same fit using also the scale dependent spectral index
given in (\ref{runpred}). In this case we obtain an allowed region in the
plane $s$ vs $c$, shown in Figure 2. In the same figure it is also
displayed the most reasonable region of values for the parameters,
from theoretical arguments (see \cite{lc00} for a full discussion).

In Table 3 are again summarized the results of the fit in this case:
we have fixed the value of $c$, so that the number of degrees of
freedom is the same as before. The value of $n$ in the table, refers
to a scale corresponding to the COBE measurement. Since $c$ is 
positive, smaller scales present larger spectral index 
$n > n_{COBE}$, e.g. $n(k_8) = 1.009$. 
Note that the value of $\chi^2$ is in this case larger:
the minimum lies on the $c=0$ line and favours no scale--dependence.
Note anyway that along the quasi degenerate $s-c$ direction, 
corresponding to small $n_{COBE}$, values of $|c| \leq 0.1-0.2$ 
are within the 70\% CL contour.

\begin{figure}
\begin{center}
\includegraphics[width=.65\textwidth]{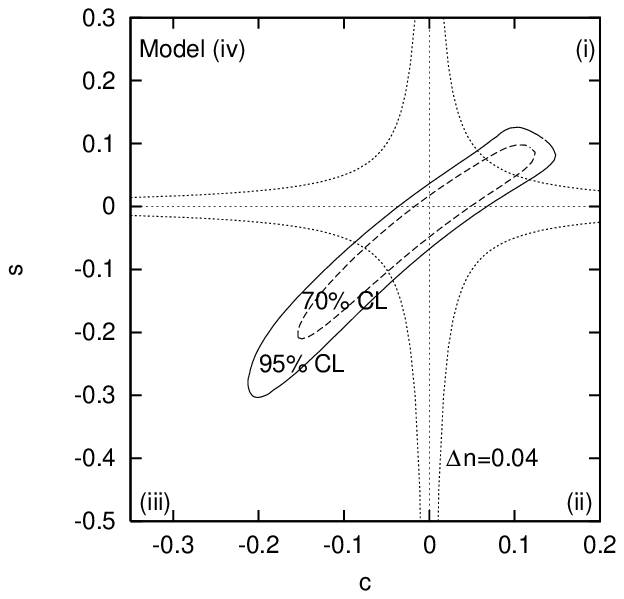}
\includegraphics[width=.6\textwidth]{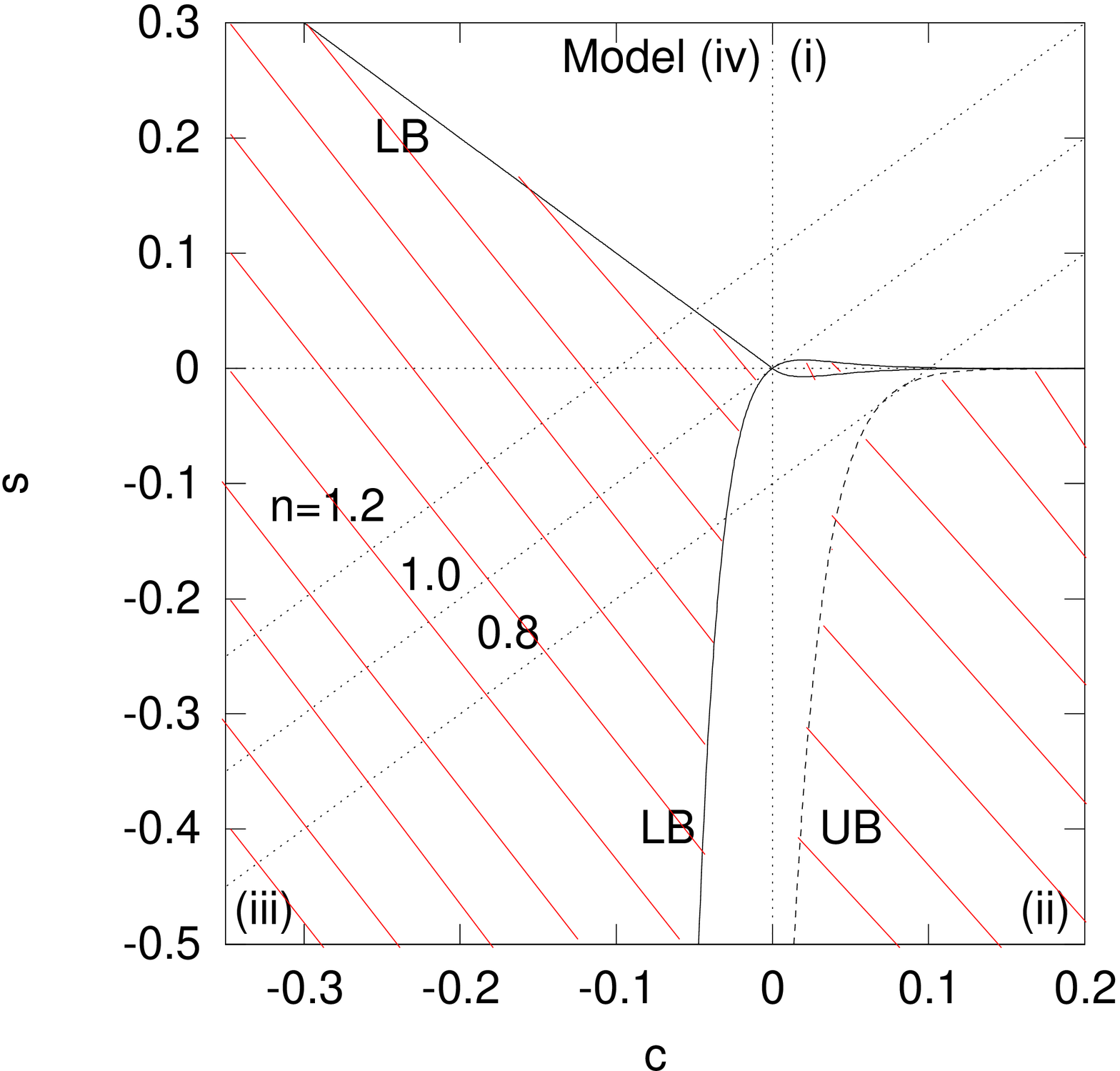}
\end{center}
\caption[]{
The  parameter space for the running mass model. As discussed in \cite{cl00}
the model comes in four versions, corresponding to the four quadrants
of the parameter space.
In panel (a), we show the region allowed by observation,
in  the case that reionization occurs when  $f\sim 1$;
the scale-dependence of the prediction for $n$ is also displayed in 
this panel by the branches of the hyperbola 
$ 8 sc = \Delta n \equiv n_8-n_{COBE} $, 
for the reference value $\Delta n=0.04$.
In panel (b),  the straight lines correspond to $n_{COBE}=1.2, 1.0$ 
and $0.8$, and the shaded region is disfavoured on  theoretical grounds
}
\label{s-c-f1}
\end{figure}

\begin{table}
\caption{Fit  of the $\Lambda$CDM model to presently available data
for the running mass models. We have fixed $c=0.1$ and so the free
parameters are $n_{COBE}=1+2(s-c)$, and the next three 
quantities in the Table. Reionization is taken to occur 
 when a fraction $f=10^{-2.2}$ of matter has collapsed, as done 
previously (the corresponding redshift at best fit is $z_R=21$.)
 Every quantity except $n_{COBE}$ is 
a data point, with the value and uncertainty listed in
the first two rows. The result of the  least-squares fit is given in the
lines three to five.  All uncertainties are at the nominal 1-$\sigma$ level. 
The total $\chi^2$ is 8.4 with three degrees of freedom }
\begin{center}
\renewcommand{\arraystretch}{1.4}
\setlength\tabcolsep{7pt}
\begin{tabular}{@{}lllllllll}
\hline\noalign{\smallskip}
& $n_{\makebox{\tiny COBE}}$ & $\Omega_b h^2$ & $\Omega_0$ & $h$ 
&$\widetilde \Gamma$ & $\widetilde \sigma_8$ & $\widetilde C_\ell^{1st}$ &
${\widetilde C_\ell^{2nd}\over \widetilde C_\ell^{1st}}$\\
\noalign{\smallskip}
\hline\noalign{\smallskip}
data & --- & 0.019 & 0.35 & 0.65 &
 0.23 & 0.56 & $74.0\, \umu {\mathrm K}$ & 0.38 \\
err. & --- & 0.002 & 0.075 & 0.075  & 0.035 & 0.059 & $5.0\, \umu {\mathrm K}$
& $.06$
 \\
fit & 0.94 & $0.021$ & $0.40$ & $0.59$
 & 0.19  & 0.53 & $67.6\, \umu {\mathrm K}$ & 0.49\\
err. & 0.04  & 0.002 & 0.05 & 0.05  & --- & --- & --- & --- \\
$\chi^2$ & --- & 0.9  & 0.4 & 0.6 &
 $1.2$ & $0.2$ & $1.6$ & 3.5 \\
\noalign{\smallskip}
\hline
\end{tabular}
\end{center}
\label{table3}
\end{table}

\newpage

\section{Conclusion}

We have  presented a fit of the $\Lambda$CDM model to a
global data set, assuming that a gaussian  primordial curvature perturbation
is the only one. We focused  on the spectral index $n$, specifying the shape
of the curvature perturbation, considering separately the case of 
a practically scale-independent spectral index, and the scale-dependent
spectral index predicted by running mass inflation models.
In contrast with other groups, we calculate the reionization epoch 
within the model on the assumption that it corresponds to the 
epoch when some fraction $f$ of the matter collapses, the results
being only mildly dependent on  $f$ in the reasonable range $f\gsim 10^{-4}$.

For the scale-independent case, we obtain a spectral index $n=0.99\pm 0.05$
for an intermediate value of $f$. The result for all $f$ is given in 
Figure 1. We stress that the best fit result strongly depends on 
the procedure used to treat the reionization redshift $z_R$.

In the case of running mass models, the scale-dependent spectral index
depends on parameters $s$ and $c$, the latter being related to the
inflaton coupling which produces the running. We have delineated
the allowed region in the $s$-$c$ plane. An unsuppressed coupling
$c\sim 0.1$ is allowed by the data, leading to a noticeable
scale-dependence of the spectral index. The  fit with $c=0.1$
 is less good than with a scale-independent spectral index,
but still acceptable.

\section*{Acknowledgments}

I am very grateful and indebted to David H. Lyth, with whom this work has
been done. I would like to thank prof. H.V. Klapdor-Kleingrothaus and the
Organizing Committee of DARK2000 for their kind invitation to participate 
to this very interesting workshop and for partial financial support.

%

\end{document}